\documentclass[aps,prd,twocolumn,showpacs,superscriptaddress,letterpaper,preprintnumbers,asmmath,amssymb]{revtex4}

\usepackage{graphicx}
\usepackage{dcolumn}
\usepackage{bm}

\RequirePackage{xspace}
\usepackage{relsize}

\def\Kbar  {\kern 0.2em\overline{\kern -0.2em K}{}\xspace}

\def\Bbar    {\kern 0.18em\overline{\kern -0.18em B}{}\xspace}

\def\Qbar    {\kern 0.08em\overline{\kern -0.08em Q}{}\xspace}

\newcommand{\mev}{\ensuremath{\mathrm{\,Me\kern -0.1em V}}\xspace}
\newcommand{\mevc}{\ensuremath{{\mathrm{\,Me\kern -0.1em V\!/}c}}\xspace}
\newcommand{\mevcc}{\ensuremath{{\mathrm{\,Me\kern -0.1em V\!/}c^2}}\xspace}
\newcommand{\gev}{\ensuremath{\mathrm{\,Ge\kern -0.1em V}}\xspace}
\newcommand{\gevc}{\ensuremath{{\mathrm{\,Ge\kern -0.1em V\!/}c}}\xspace}
\newcommand{\gevcnospace}{\ensuremath{{\mathrm{\,Ge\kern -0.1em V\!/}c}}}
\newcommand{\gevcc}{\ensuremath{{\mathrm{\,Ge\kern -0.1em V\!/}c^2}}\xspace}
\newcommand{\be}{\begin{equation}}
\newcommand{\ee}{\end{equation}}
\newcommand{\benn}{\begin{equation*}}
\newcommand{\eenn}{\end{equation*}}
\newcommand{\bea}{\begin{eqnarray}}
\newcommand{\eea}{\end{eqnarray}}

\begin{document}


\preprint{UCI-TR-2010-13}

\title{ Tevatron Discovery Potential for Fourth Generation Neutrinos:
  Dirac, Majorana and Everything in Between}

\author{
 A.~Rajaraman and D.~Whiteson\\
{\small {\it Dept. of Physics and Astronomy, University of California, Irvine, Irvine, California 92697}}}

\date{\today}

\begin{abstract}
We analyze the power of the Tevatron dataset to exclude or discover
fourth generation neutrinos. In a general  framework, one can
have mixed
left- and right-handed neutrinos, with Dirac and Majorana neutrinos as
extreme cases.  We demonstrate that a single Tevatron experiment can
make powerful statements across the entire mixing space,  extending
LEP's mass limits of 60-80 GeV up to 150-175 GeV, depending on the mixing.
\end{abstract}


\maketitle

\section{Introduction}
A simple and well-motivated possibility for new physics is a fourth
generation of quarks and leptons.   Contrary to some previous
arguments~\cite{Amsler:2008zzb}, such an extension of the standard
model is entirely compatible with precision electro-weak
data~\cite{He:2001tp,Kribs:2007nz}.

Direct searches have been performed at the Tevatron for heavy fourth generation quarks
($t'$ and $b'$), placing bounds of
335 GeV~\cite{Cox:2009mx} and 338 GeV~\cite{Aaltonen:2009nr},
respectively, on their masses. The LHC will be able to discover or
definitively exclude the existence of a fourth generation of quarks up to ~1
TeV \cite{Ozcan:2008zz,Cakir:2008su,Holdom:2007ap,Han:2006ip,Atre:2009rg,delAguila:2008ir,delAguila:2007em,Hung:2007ak,CuhadarDonszelmann:2008jp,Holdom:2009rf}.

The lepton sector, though it has great promise, has not been studied in such
detail. A fourth generation of leptons may even
be lighter than the fourth generation quarks, in analogy to the first
three generations. In particular, the neutrino may well be the
lightest new particle. Furthermore, the lepton sector is expected to
be extremely rich, as the relatively high mass scale
 for the neutrino required by electroweak measurements suggests that
 the right handed neutrino of the fourth generation is not very
 massive.  The leptonic sector therefore has two neutrino states in addition to a charged lepton.

For these reasons, it is of great interest to search for potential signals of
this leptonic sector at colliders. The signals, however, are very
dependent on the precise spectrum and the three mass parameters
complicate the analysis.  We first consider the simplest case, the limit
in which the lepton and
one of the neutrinos is very heavy, and the theory reduces to a single Majorana neutrino.
 LEP2 has placed constraints on fourth generation
neutrino masses
 in this limit~\cite{Achard:2001qw}, depending on the decay mode of
 the neutrino.
If the neutrino decay mode were $N\rightarrow eW$, the neutrino mass is constrained
to be larger than 90.7 GeV, with
corresponding bounds at  89.5 and 80.5 GeV respectively for
$\mu W$ and $\tau W$ decay modes.

The more general case has
two relatively light neutrinos of comparable mass. (We shall assume in this article that the lepton is significantly more massive
than the two neutrinos and decouples;
we leave its inclusion for future work.)
This limit was considered in
a recent paper~\cite{Carpenter:2010dt} which reanalyzed the  bounds from the LEP2 analysis on this parameter space.
It was shown in \cite{Carpenter:2010dt} that
the LEP bounds on the neutrino mass should be weakened significantly as compared to
the single neutrino case; for example, a neutrino decaying to $\tau$W could be as light as 62.1 GeV,
as compared to the original bound of 80.5 GeV in the limit where only one neutrino is light
(bounds were reduced to about 80 GeV  for
eW and $\mu$W   decay modes). There is therefore a large parameter
space newly opened
for analysis.

This article addresses the open question of whether the Tevatron can improve these
bounds in the general two-neutrino case. The Tevatron has not performed an analysis; however, a 
sensitivity study~\cite{{rajwhiteson}} indicated that the Tevatron could significantly 
improve the mass bounds on the neutrino in the one-neutrino limit.
 For this limit, the sensitivity study showed that, at least for the $\mu$W decay mode, the Tevatron
could expect to dramatically improve the mass bounds to 175 GeV.  This suggests that
for the  two-neutrino parameter space, the Tevatron will again have significant reach.

 In this article, we present a sensitivity study for the
Tevatron in the more general two-neutrino mixing space and show that in fact the neutrino mass bounds can be significantly
improved. We note that previous studies have explored the possibility of searching for
neutrinos at the LHC~\cite{CuhadarDonszelmann:2008jp} and a future ILC~\cite{Ciftci:2005gf}; however, a Tevatron sensitivity study has not been performed.

We begin by reviewing the theory of the two-neutrino system, and discussing the production and decay
of these particles at hadron colliders. We then calculate experimental
sensitivity for a selection of same-charge leptons, using Monte Carlo
simulation. Finally we discuss our results and conclude.

\section{Production and Decay of Fourth Generation Neutrinos}
The mass term for the two-neutrino system can be written as
 \cite{Carpenter:2010dt,Katsuki:1994as}
\begin{eqnarray}
L_m=-{1\over 2}\overline{(Q_R^c
N_R^c)}\left(\begin{array}{cc}0&m_D\\m_d&M\end{array}\right)
\left(\begin{array}{c}Q_R\\ N_R\end{array}\right)+h.c.
\end{eqnarray}

The mass eigenstates are
 \bea
 N_1=c_\theta N_L^c+s_\theta N_R+c_\theta N_L+s_\theta N_R^c
 \\
 N_2=-is_\theta N_L^c+ic_\theta N_R+is_\theta N_L-ic_\theta N_R^c
 \eea
with corresponding eigenvalues
 \bea
 M_1=-(M/2)+ \sqrt{m_D^2+{M^2/4}}\\
M_2=(M/2)+ \sqrt{m_D^2+{M^2/4}}
\eea
Here $\psi^c=-i\gamma^2\psi^*$ and $Q_R=N_L^c$.
The mixing angle is related to the masses by the relation
\bea
 \cos^2\theta=
 {M_2\over M_2+M_1}
 \eea
 and varies over the range ${1\over 2}\le  \cos^2\theta\le 1$.

The  neutrinos couple to the gauge bosons through the interaction term
$L=gW_\mu^+ J^{\mu +}
 +gW_\mu^- J^{\mu -}+gZ_\mu J^{\mu}$ where
\bea
 J^{\mu }
= {1\over 2\cos\theta_W}(-c^2_\theta\bar N_1\gamma^\mu \gamma^5N_1
-2is_\theta c_\theta\bar N_1\gamma^\mu N_2\nonumber\\
-s^2_\theta\bar N_2 \gamma^\mu\gamma^5N_2))
\\
J^{\mu +}=
c_i\overline{(c_\theta N_1-i s_\theta N_2)}\gamma^\mu l^i_L~~~~~~~~~~~~~~~~~~~
\eea
where $c_i$ are analogous to the CKM matrix elements.

At colliders, the neutrinos can be produced  either through the process $q\bar{q}'\rightarrow W^{\pm}\rightarrow N_il^{\pm}$,  where one of the fourth generation
neutrinos is produced in association with a light charged lepton, or through the process
$q\bar{q}\rightarrow Z\rightarrow N_iN_j$, where two heavy neutrinos are produced. There are many papers studying the reach of the Tevatron and LHC to the $W$ process
\cite{Han:2006ip,Atre:2009rg,delAguila:2008ir,delAguila:2007em}.
This is  because the $W$ production has a higher cross section at hadron colliders, and
is expected to dominate.
Furthermore, the  mass reach is enhanced because only one heavy particle is produced in this process.
On the other hand, the  cross-section for the first process depends on the values of $c_i$, the parameters that control the
mixing between the fourth generation with the first three generations.
These constants are not theoretically calculable, but precision measurements
suggest that this angle is small. If the mixing angle is less than
about $10^{-6}$, the neutrino production rates in this channel are suppressed enough that
they are unobservable at colliders~\cite{Han:2006ip}.   The rate of
heavy neutrino pair-production via a $Z$ boson, however, does not
depend on the mixing parameter. We will assume that we are in the regime where this mixing angle is small, and production through
an $s$-channel $Z$ is the dominant production mechanism.

\begin{figure}[t]
\begin{center}
\includegraphics[width=0.48\linewidth]{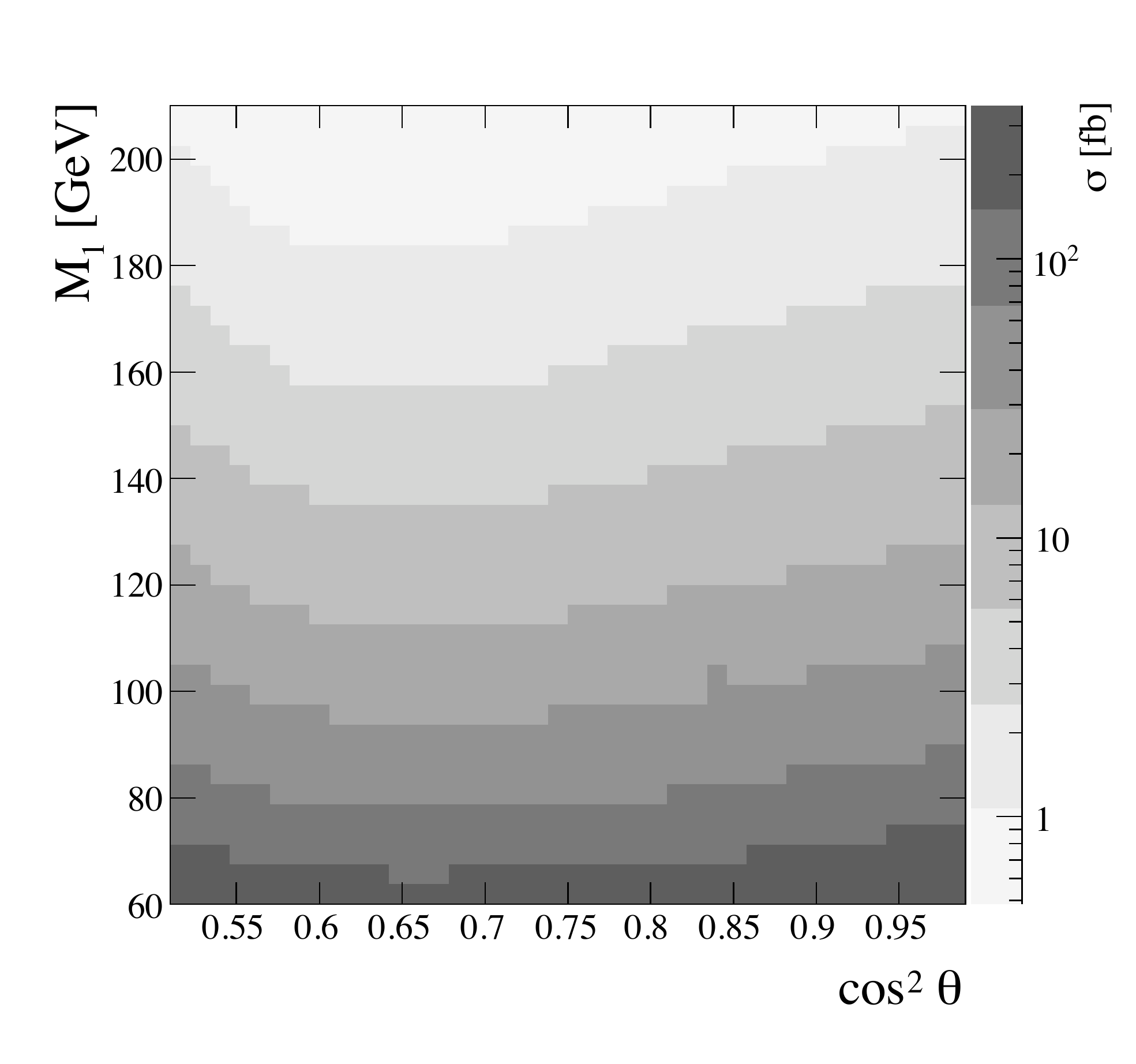}
\includegraphics[width=0.48\linewidth]{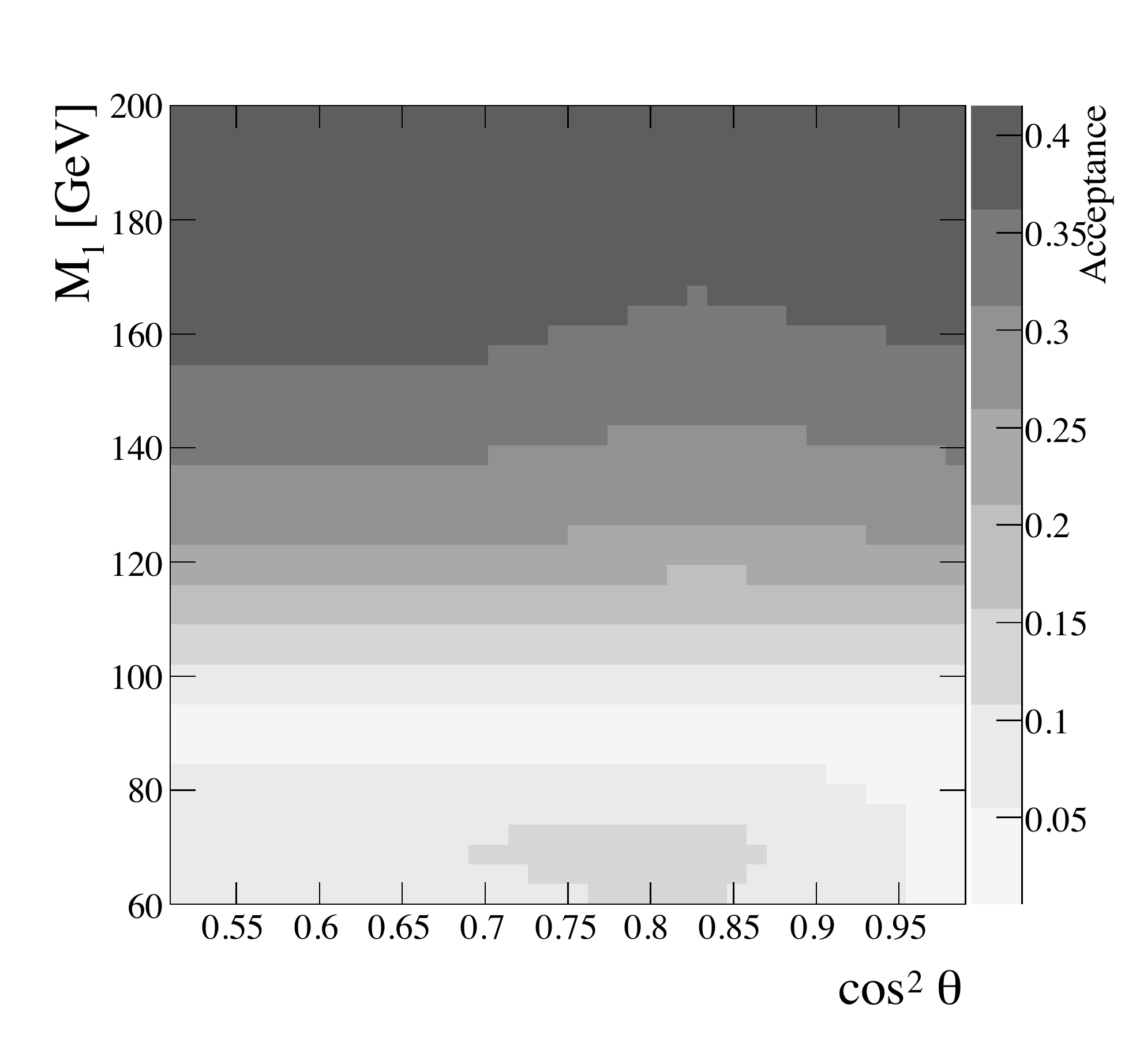}
\caption{ Left, cross-section for $p\bar{p} \rightarrow Z \rightarrow
  NN$ where $NN = N_1N_1 + N_1N_2 + N_2N_2$ as a function of the $N_1$
  mass ($M_1$) and the mixing angle $\theta$. Right, acceptance of our
  selection, averaged over the $N_1N_1, N_1N_2, N_2N_2$ modes if BR($N \rightarrow \mu W)=100$\%.}
\label{fig:xa}
\end{center}
\end{figure}

The decay time of the neutrinos also depend on the unknown $c_i$, but $N_1$ always decays to $lW$, where $l$
is a lepton of the first  three generations. We will assume that the 
decay happens promptly and that the neutrino does not escape or leave 
displaced vertices; this will happen unless the mixing angle is 
extremely tiny~\cite{Hung:2007ak}. $N_2$  decays to $lW$ or $N_1Z$; the $lW$ channel is
suppressed by the small $c_i$, and the $N_1Z$ channel will dominate
except if the mass difference is very small. We will assume that the mass difference
is at least 1 GeV, and we assume that the CKM factor is so small that the $N_1Z$  decay
always dominates in this range.

Note that in the exact Dirac limit, $N_2$  must decay to $lW$.
In this limit, the different contributions to same sign dilepton
production cancel.
This is expected since the Dirac fermion conserves fermion number.
However, since we are assuming that  $N_2$ always decays
to $N_1 Z$ (i.e. we do not take the exact Dirac limit), there is no
interference amplitude, giving same-sign
dilepton decays in the entire mixing space.

We therefore consider the processes
\bea
pp\rightarrow Z\rightarrow N_1N_1\rightarrow lWlW~~~~
\\
pp\rightarrow Z\rightarrow N_1N_2\rightarrow lWlWZ~~
\\
pp\rightarrow Z\rightarrow N_2N_2\rightarrow lWlWZZ
\eea
In each case, half the decays have same sign leptons and
correspondingly same sign $W$s.


  Figure~\ref{fig:xa} shows the total
cross-section for all three processes as a function of $N_1$ mass and
mixing angle $\theta$.    Decays via a Higgs boson were also considered
($h\rightarrow N_1N_1,h\rightarrow N_2N_2$), but the large Higgs mass
required to pair produce the heavy neutrinos
makes the Higgs
contribution small as compared to the production via $Z$.

\section{Experimental Sensitivity}

\begin{figure}[t]
\begin{center}
\includegraphics[width=0.48\linewidth]{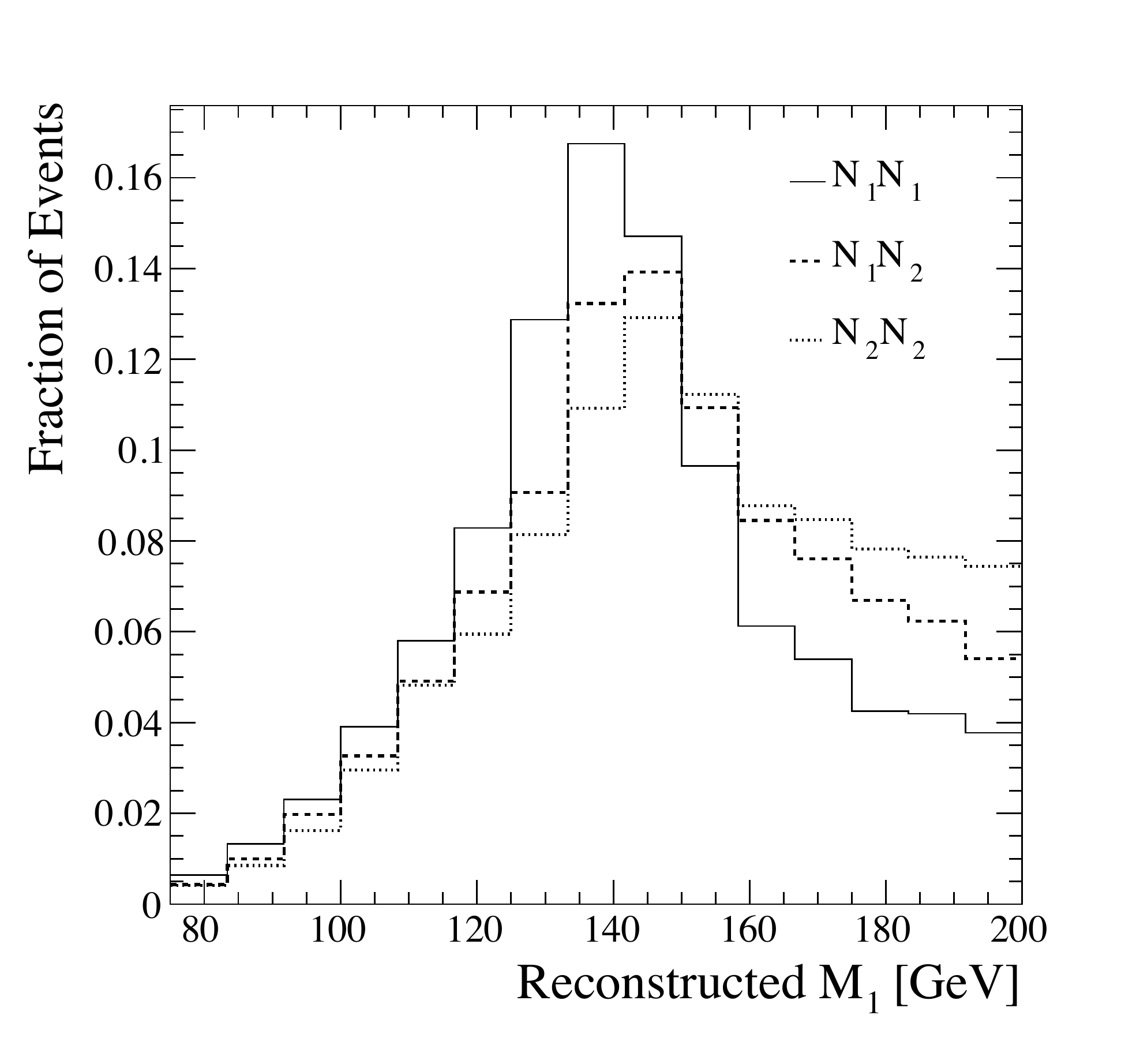}
\includegraphics[width=0.48\linewidth]{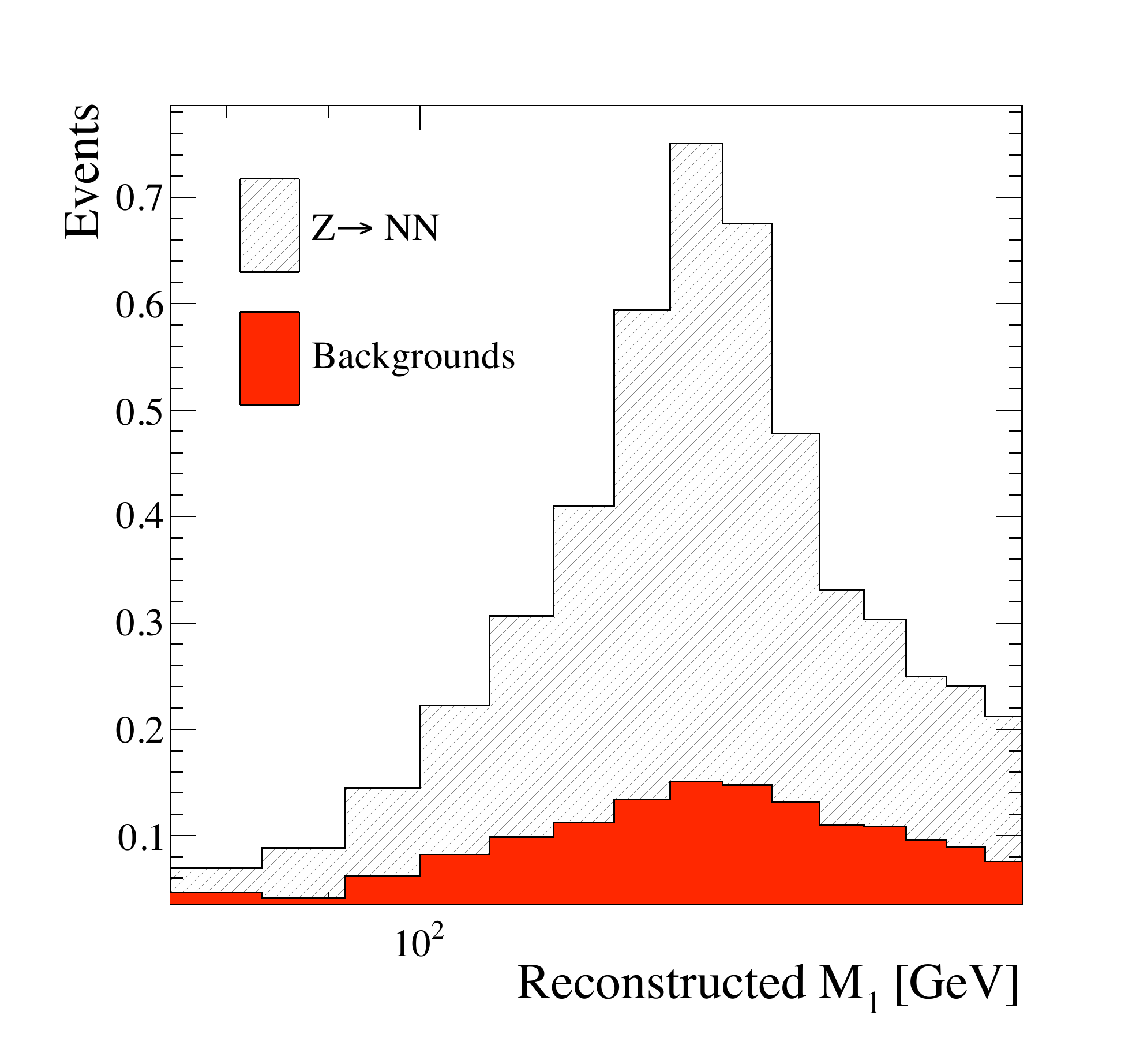}
\caption{ For $M_1 = 150$ GeV, and cos$^2\theta$=0.67. Left,
  reconstructed $M_1$ in selected events with for $N_1N_1,
  N_1N_2, N_2N_2$ modes. Other masses and mixing angles are
  qualitatively similar. Right, expected signal and backgrounds in
  5/fb of CDF data.}
\label{fig:reco}
\end{center}
\end{figure}

We study the most sensitive region, where BR($N \rightarrow \mu
W)=100$\%. Following ~\cite{rajwhiteson}, we select events with the $\ell^{\pm}\ell^{\pm}jj$ signature:
\begin{itemize}
\item two like-signed reconstructed muons each with $p_T > 20$ GeV
  and $|\eta|<2.0$
\item at least two reconstructed jets, each with $p_T > 15$ GeV
  and $|\eta|<2.5$
\end{itemize}
The efficiency of the selection is shown in Figure~\ref{fig:xa} and
is a strong function of $N_1$ mass; when $M_{1} - m_{W}$ is small, the lepton from the $N_1\rightarrow lW$ decay has small
transverse momentum and is difficult to reconstruct.

We reconstruct one or two $N_1$ masses in each event using
all $m_{jjl}$ combinations where $m_{jj}$ is consistent with a
hadronic $W$ decay.  Figure~\ref{fig:reco} shows the reconstructed
$N_1$ mass shape for the  $N_1N_1,  N_1N_2,$ and $N_2N_2$ modes.

At CDF, the largest backgrounds to the  $\ell^{\pm}\ell^{\pm}jj$ signature come
from $W\gamma$ or $WZ$ production or misidentified leptons~\cite{cdf1fbprl} either from
semi-leptonic $t\bar{t}$ decays or direct $W+$ jets production.  As in
~\cite{rajwhiteson}, we
extrapolate the number of expected backgrounds events 
in 1 fb$^{-1}$~\cite{cdf1fbprl} to a dataset with 5 fb$^{-1}$, use {\sc madgraph}~\cite{madgraph} to 
model the kinematics
of the events, {\sc pythia}~\cite{pythia} for showering and a version of
{\sc pgs}~\cite{pgs} tuned to describe the performance of the CDFII
detector.  Figure~\ref{fig:reco} shows the expected signal and
backgrounds in 5 fb$^{-1}$ of CDF data.
We perform a binned likelihood fit in the reconstructed $N_1$
mass, and use the unified ordering scheme~\cite{feldcous} to calculate
median expected limits from frequentist intervals.


\begin{figure}[t]
\begin{center}
\includegraphics[width=0.8\linewidth]{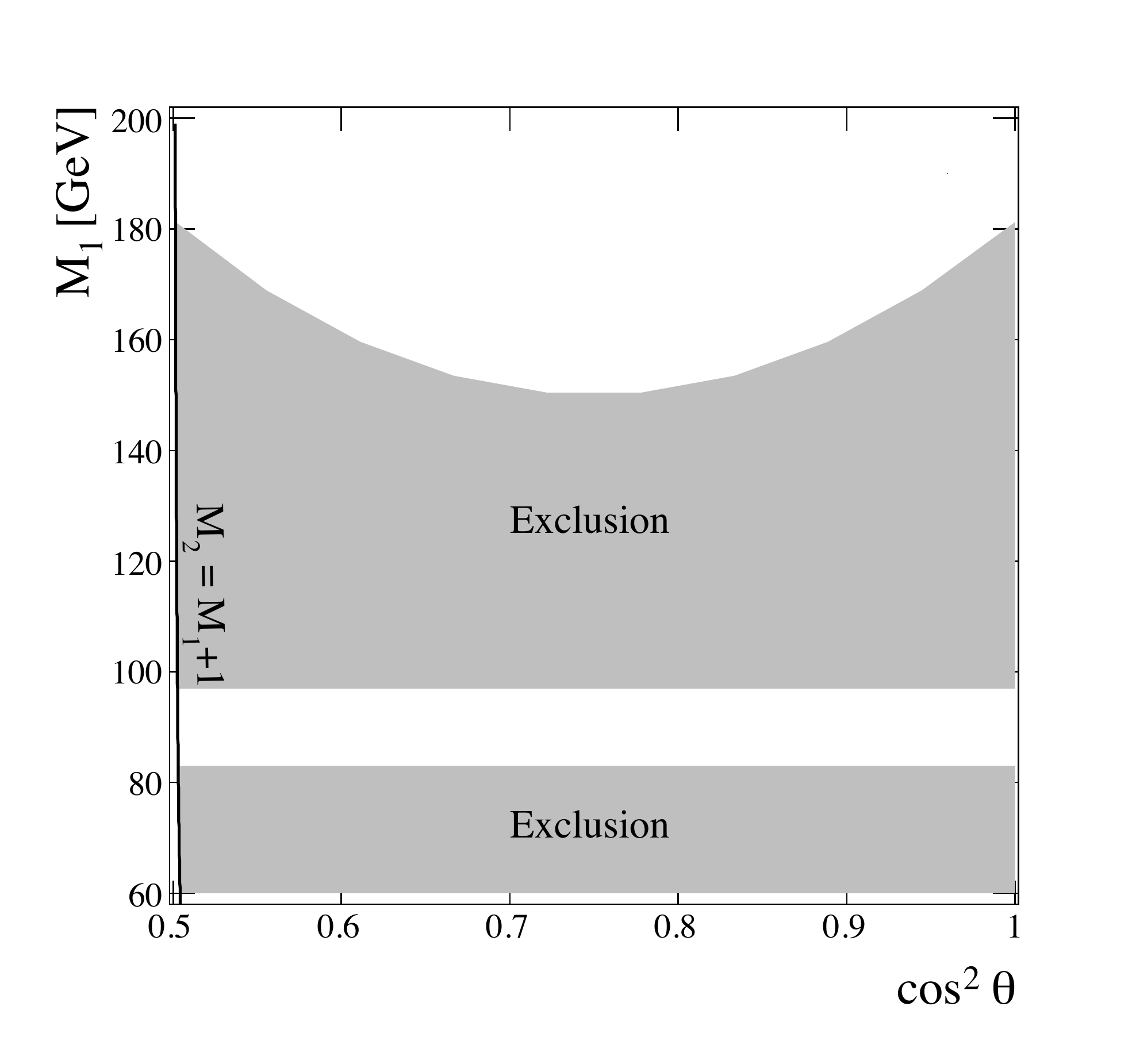}
\caption{ Expected 95\% C.L. exclusion regions for 5 fb$^{-1}$ of CDF data, as a function of the $N_1$
  mass and the mixing angle $\theta$.}
\label{fig:sens}
\end{center}
\end{figure}


 We present the expected Tevatron constraints in the $(M_1,\cos^2\theta)$ plane in Fig {\ref{fig:sens}}.
  The shaded regions show regions that can be excluded by a Tevatron
search, assuming BR($N \rightarrow \mu W)=100$\%.

The shape of the constraints can be understood as follows.
In the limit  $\cos^2\theta= 1$, $M_2$ is
 infinite, and we return to the one-neutrino case.
 As $\cos^2\theta$ decreases,
 the mixing angle between the Z and the lighter neutrino
is reduced, leading to a smaller cross-section and therefore a weakening of the bounds. On the other hand,
as $\cos^2\theta$  decreases, the heavier neutrino mass is also reduced (when $\cos^2\theta= 1/2$, the neutrinos are degenerate); as a 
consequence, as we reduce $\cos^2\theta$ from
1 to 1/2, the heavier neutrino eventually becomes light enough that it is accessible
to production. At this point, the bounds again improve.
The mass exclusion therefore weakens and then strengthens as a function of $\cos^2\theta$.
Furthermore, at  lower values of $M_1$, there is a gap which is not excluded by our study. This occurs because the leptons from $N\rightarrow lW$ become
very soft, and fall below the $p_T$ cut.

For $\cos^2\theta$ very close to 1/2, we approach the Dirac limit. In this limit, the decay $N_2\rightarrow
N_1 Z$ may open up (depending on the mixing between the fourth and the other generations).
We therefore impose the condition $M_2>M_1+1$. There is thus a  narrow strip on the left of the plot which is
excluded from our analysis.


Our analysis has assumed that the neutrino only decays to $\mu$W.
If the neutrino decays to eW or $\tau$W as well, the limits will be degraded.
This was analyzed for the one-neutrino case in
\cite{rajwhiteson} and it was shown how the limits are reduced. Since
we describe the identical selection here, our limits will degrade in
the same manner.

\section{Conclusions}

We have shown that the Tevatron has a significant reach to fourth generation neutrinos, and
can place significant constraints on the general two-neutrino parameter space.
Quantitatively, if the $N_1,N_2$ do not exist and no excess is
seen, CDF can exclude the existence of $N_1,N_2$ up to 150-170 GeV,
depending on the mixing angle, with the exception of a band between
83-97 GeV, where the acceptance is very small due to the softness of
the produced lepton.  This makes a strong case to  search for
potential signals of these neutrinos in the current dataset.

These results leave open several interesting directions for future
research. Perhaps the most difficult and important is to cover the gap where the
neutrino mass is close to the $W$ mass by understanding the
backgrounds to very soft leptons.  In addition, we hope to include the charged
leptons in future studies.

\section{Acknowledgements}

We acknowledge discussions with L.~Carpenter and T.~ Tait. A.R. is
supported in part by NSF Grant No. PHY-0653656. D.W. is supported in part by the U.S. Dept. of Energy.




\begin{thebibliography}{99}



\bibitem{Amsler:2008zzb}
  C.~Amsler {\it et al.}  [Particle Data Group],
  Phys.\ Lett.\  B {\bf 667}, 1 (2008).

\bibitem{He:2001tp}
  H.~-J.~He, N.~Polonsky, S.~-f.~Su,
  Phys.\ Rev.\  {\bf D64}, 053004 (2001).
  [hep-ph/0102144].

\bibitem{Kribs:2007nz}
  G.~D.~Kribs, T.~Plehn, M.~Spannowsky and T.~M.~P.~Tait,
  Phys.\ Rev.\  D {\bf 76}, 075016 (2007)
  [arXiv:0706.3718 [hep-ph]].




\bibitem{Cox:2009mx}
  D.~Cox  [CDF Collaboration and for the CDF Collaboration],
  arXiv:0910.3279 [hep-ex].

\bibitem{Aaltonen:2009nr}
  T.~Aaltonen {\it et al.}  [The CDF Collaboration],
  arXiv:0912.1057 [hep-ex].













\bibitem{Ozcan:2008zz}
  V.~E.~Ozcan, S.~Sultansoy and G.~Unel,
  Eur.\ Phys.\ J.\  C {\bf 57}, 621 (2008).
\bibitem{Cakir:2008su}
  O.~Cakir, H.~Duran Yildiz, R.~Mehdiyev and I.~Turk Cakir,
  Eur.\ Phys.\ J.\  C {\bf 56}, 537 (2008)
  [arXiv:0801.0236 [hep-ph]].
\bibitem{Holdom:2007ap}
  B.~Holdom,
  JHEP {\bf 0708}, 069 (2007)
  [arXiv:0705.1736 [hep-ph]].


\bibitem{Han:2006ip}
  T.~Han and B.~Zhang,
  Phys.\ Rev.\ Lett.\  {\bf 97}, 171804 (2006)
  [arXiv:hep-ph/0604064].
\bibitem{Atre:2009rg}
  A.~Atre, T.~Han, S.~Pascoli and B.~Zhang,
  JHEP {\bf 0905}, 030 (2009)
  [arXiv:0901.3589 [hep-ph]].

\bibitem{delAguila:2008ir}
  F.~del Aguila, S.~Bar-Shalom, A.~Soni and J.~Wudka,
  Phys.\ Lett.\  B {\bf 670}, 399 (2009)
  [arXiv:0806.0876 [hep-ph]].

\bibitem{delAguila:2007em}
  F.~del Aguila, J.~A.~Aguilar-Saavedra and R.~Pittau,
  JHEP {\bf 0710}, 047 (2007)
  [arXiv:hep-ph/0703261].



\bibitem{Hung:2007ak}
  P.~Q.~Hung and M.~Sher,
  Phys.\ Rev.\  D {\bf 77}, 037302 (2008)
  [arXiv:0711.4353 [hep-ph]].

\bibitem{CuhadarDonszelmann:2008jp}
  T.~Cuhadar-Donszelmann, M.~K.~Unel, V.~E.~Ozcan, S.~Sultansoy and G.~Unel,
  JHEP {\bf 0810}, 074 (2008)
  [arXiv:0806.4003 [hep-ph]].



  \bibitem{Holdom:2009rf}
  B.~Holdom, W.~S.~Hou, T.~Hurth, M.~L.~Mangano, S.~Sultansoy and G.~Unel,
  PMC Phys.\  A {\bf 3}, 4 (2009)
  [arXiv:0904.4698 [hep-ph]].



  \bibitem{Achard:2001qw}
  P.~Achard {\it et al.}  [L3 Collaboration],
  Phys.\ Lett.\  B {\bf 517}, 75 (2001)
  [arXiv:hep-ex/0107015].


\bibitem{Carpenter:2010dt}
  L.~M.~Carpenter and A.~Rajaraman,
  [arXiv:1005.0628 [hep-ph]].


\bibitem{rajwhiteson}
  A. Rajaraman and D. Whiteson
  Phys.\ Rev.\  D {\bf 81}, 071301 (2010).
  
\bibitem{Ciftci:2005gf}
  A.~K.~Ciftci, R.~Ciftci and S.~Sultansoy,
  Phys.\ Rev.\  D {\bf 72}, 053006 (2005)
  [arXiv:hep-ph/0505120].

\bibitem{Katsuki:1994as}
  Y.~Katsuki, M.~Marui, R.~Najima {\it et al.},
  Phys.\ Lett.\  {\bf B354}, 363-370 (1995).
  [hep-ph/9501236].


\bibitem{cdf1fbprl} CDF Collaboration, Phys. Rev. Lett. {\bf 98}, 221803 (2007)

\bibitem{madgraph}J.~Alwall {\it et al.}, JHEP {\bf 0709} 028 (2007)

\bibitem{pythia}  T.~Sj\"ostrand {\it et al.},
  Comput. Phys. Commun. {\bf 238} 135 (2001).

\bibitem{pgs} M.~Carena {\it et al.} arXiv:hep-ph/0010338v2

\bibitem{feldcous} G.~J.~Feldman and R.~D.~Cousins, Phys. Rev. D {\bf 57}, 3873 (1998).


\end{thebibliography}
\end{document}